\documentclass[12pt,thmsa]{article}

\usepackage{cite}
\usepackage{amsmath}

\begin{document}

\title{Perturbation approaches and Taylor series}
\author{Francisco M. Fern\'{a}ndez \\
INIFTA (UNLP, CCT La Plata--CONICET), Divisi\'{o}n Qu\'{i}mica Te\'{o}rica,\\
Diag. 113 y 64 (S/N), Sucursal 4, Casilla de Correo 16,\\
1900 La Plata, Argentina \\
e--mail: fernande@quimica.unlp.edu.ar}
\date{}
\maketitle

\begin{abstract}
We comment on the new trend in mathematical physics that consists
of obtaining Taylor series for fabricated linear and nonlinear
unphysical models by means of homotopy perturbation method (HPM),
homotopy analysis method (HAM) and Adomian decomposition method
(ADM). As an illustrative example we choose a recent application
of the HPM to a dynamic system of anisotropic elasticity.
\end{abstract}

\section{Introduction}

\label{sec:intro} In the last years there has been great interest in the
application of approximate variational and perturbation methods that lead to
power--series solutions for linear and nonlinear problems in mathematical
physics\cite{SNH07,CHA07,CH07a,EG07,SNH08,SNH08b,CH08,
M08,OA08,RAH08,SG08,ZLL08,KY09}. For example, Chowdhury and Hashim\cite
{CH07a} applied the powerful homotopy perturbation method (HPM) to obtain
the Taylor series about $x=0$ of the functions $y(x)=e^{x^{2}}$, $%
y(x)=1-x^{3}/3!$, $y(x)=\sin (x)/x$, and $y(x)=x^{2}+x^{8}/72$. By means of
the same method Chowdhury et al\cite{CHA07} derived the Taylor series about $%
t=0$ for the solutions of the simplest population models. Bataineh et al\cite
{SNH07} went a step further and resorted to the even more powerful homotopy
analysis method (HAM) and calculated the Taylor expansions about $x=0$ of
the following two--variable functions: $y(x,t)=e^{x^{2}+\sin t}$, $%
y(x,t)=x^{2}+e^{x^{2}+t}$, $y(x,t)=x^{3}+e^{x^{2}-t}$, $%
y(x,t)=t^{2}+e^{x^{3}}$, and most impressive: $y(x,t)=-2\ln (1+tx^{2})$ and $%
y(x,t)=e^{-tx^{2}}$. Zhang et al\cite{ZLL08} dared to meddle with functions
as complex as $u(x,t)=-2\sec h^{2}[(x-2t)/2]$, and $u(x,t)=-(15/8)\sec
h^{2}[(x-5t/2)]$. Because of the inherent difficulty in such functions the
authors pushed the HPM calculation just to first order in $t$ (I mean only
the linear term of the Taylor series about $t=0$). Bataineh et al\cite
{SNH08b} showed great insight and modified HAM to produce MHAM and found the
Taylor expansions of the functions $u(t)=t^{2}-t^{3}$ and $u(t)=1+t^{2}/16$.
Sadighi and Ganji\cite{SG08} calculated the Taylor expansions about $t=0$ of
$u(x,t)=1+\cosh (2x)e^{-4it}$ and $u(x,t)=e^{3i(x+3t)}$ by means of HPM and
Adomian decomposition method (ADM), and verified that the results were
exactly the same!!. These authors dared to face the fact that $i^{2}=-1$. By
means of HPM Rafiq et al\cite{RAH08} also derived polynomial functions like $%
y(x)=x^{4}-x^{3}$, $y(x)=x^{2}+x^{3}$ and $y(x)=x^{2}+x^{8}/72$. \"{O}zis
and Agirseven\cite{OA08} astonished the mathematics and physics community by
expanding $u(x,t)=x^{2}e^{t}$, $u(x,y,t)=y^{2}\cosh t+x^{2}\sinh t$ (and
other such functions) about $t=0$ by means of the HPM. Bataineh et al\cite
{SNH08} used HAM to obtain expansions about $x=0$ for $w(x,t)=xe^{-t}+e^{-x}$%
, $w(x,t)=e^{x+t+t^{2}}$, $w(x,t)=e^{t+x^{2}}$ and $w(x,t)=e^{t^{2}+x^{2}}$.
They thus managed to reproduce earlier ADM results. Although the authors did
not state it explicitly in our opinion one of their achievements was to
prove that the Taylor series have exactly the same form in different world
locations.

There are many more articles where the authors apply HPM, HAM and ADM and
produce results that an undergraduate student would easily obtain by means
of a straightforward Taylor expansion of the model differential equations.
It is a new trend in mathematical physics and we have just collected some
examples published in only one journal.

In a recent paper Ko\c{c}ak and Y\i ld\i r\i m\cite{KY09} applied HPM to a
3D Green's function for the dynamic system of anisotropic elasticity. The
purpose of this article is to show the close connection between their
results and the Taylor series approach. In Sec.~\ref{sec:HPM} we introduce
the problem and outline the application of the HPM. In Sec.~\ref{sec:Taylor}
we apply the Taylor--series approach to the same problem. Finally in Sec.~%
\ref{sec:conclusions} we discuss the results and draw conclusions.

\section{The homotopy perturbation method}

\label{sec:HPM} Ko\c{c}ak and Y\i ld\i r\i m\cite{KY09} considered equations
of the form
\begin{equation}
\mathbf{\rho }\frac{\partial ^{2}\mathbf{u}(\mathbf{x},t)}{\partial t^{2}}=%
\hat{L}\mathbf{u}(\mathbf{x},t)+\mathbf{f}(\mathbf{x},t)  \label{eq:dif_eq}
\end{equation}
where $\mathbf{u},\mathbf{f\in R}^{m}$ are vector functions of time and the
spatial variables $\mathbf{x}\in R^{n}$, $\mathbf{\rho \in R}^{m\times m}$
is an invertible constant matrix and $\hat{L}$ a differential operator
independent of $t$.

They introduced a perturbation parameter $p$ into the equation (\ref
{eq:dif_eq})
\begin{equation}
\mathbf{\rho }\frac{\partial ^{2}\mathbf{u}(\mathbf{x},t)}{\partial t^{2}}%
=p\left[ \hat{L}\mathbf{u}(\mathbf{x},t)+\mathbf{f}(\mathbf{x},t)\right]
\label{eq:dif_eq_p}
\end{equation}
and solved it by means of straightforward perturbation theory (with the
fancy name of HPM)
\begin{equation}
\mathbf{u}(\mathbf{x},t)=\sum_{j=0}^{\infty }p^{j}\mathbf{u}^{(j)}(\mathbf{x}%
,t)  \label{eq:u_p_series}
\end{equation}
Thus, they obtained
\begin{eqnarray}
\frac{\partial ^{2}\mathbf{u}^{(0)}(\mathbf{x},t)}{\partial t^{2}} &=&%
\mathbf{0}  \nonumber \\
\frac{\partial ^{2}\mathbf{u}^{(j)}(\mathbf{x},t)}{\partial t^{2}} &=&%
\mathbf{\rho }^{-1}\left[ \hat{L}\mathbf{u}^{(j-1)}(\mathbf{x},t)+\mathbf{f}(%
\mathbf{x},t)\delta _{j1}\right] ,\;j=1,2,\ldots  \label{eq:HPM_eqs}
\end{eqnarray}
Those authors were mainly interested in the Green's function for the
differential equation (\ref{eq:dif_eq}) in which case $\mathbf{f}(x,t)$ is a
product of delta functions.

\section{Power series approach}

\label{sec:Taylor} Instead of applying the fashionable HPM we may try the
well--known Taylor series approach and expand the functions in equation (\ref
{eq:dif_eq}) in a Taylor series about $t=0$:
\begin{eqnarray}
\mathbf{u}(\mathbf{x},t) &=&\sum_{j=0}^{\infty }t^{j}\mathbf{u}_{j}(\mathbf{x%
})  \nonumber \\
\mathbf{f}(\mathbf{x},t) &=&\sum_{j=0}^{\infty }t^{j}\mathbf{f}_{j}(\mathbf{x%
})  \label{eq:u_t_series}
\end{eqnarray}
Thus we obtain a simple expression for the coefficients of the solution:
\begin{equation}
\mathbf{u}_{j+2}=\frac{1}{(j+1)(j+2)}\mathbf{\rho }^{-1}\left( \hat{L}%
\mathbf{u}_{j}+\mathbf{f}_{j}\right) ,\;j=0,1,\ldots  \label{eq:u_j+2}
\end{equation}
so that
\begin{eqnarray}
\mathbf{u}(\mathbf{x},t) &=&\mathbf{u}_{0}(x)+t\mathbf{u}_{1}(\mathbf{x})+%
\frac{t^{2}}{2}\mathbf{\rho }^{-1}\left( \hat{L}\mathbf{u}_{0}+\mathbf{f}%
_{0}\right) +  \nonumber \\
&&\frac{t^{3}}{6}\mathbf{\rho }^{-1}\left( \hat{L}\mathbf{u}_{1}+\mathbf{f}%
_{1}\right) +\ldots  \label{eq:u_t_series2}
\end{eqnarray}

One easily verifies that all the results shown by Ko\c{c}ak and Y\i ld\i r\i %
m\cite{KY09} are particular cases of equation (\ref{eq:u_t_series2}) with $%
\mathbf{u}_{0}=\mathbf{f}_{0}=0$ and $\hat{L}\mathbf{u}_{1}=-\mathbf{f}_{1}$
so that in every case they got the exact solution $\mathbf{u}(\mathbf{x},t)=t%
\mathbf{u}_{1}(\mathbf{x})$. As an illustrative example consider their
illustrative example\cite{KY09}
\begin{eqnarray}
\hat{L} &=&\nabla ^{2}=\frac{\partial ^{2}}{\partial x_{1}^{2}}+\frac{%
\partial ^{2}}{\partial x_{2}^{2}}  \nonumber \\
f(x,t) &=&-2t\cos ^{2}(x_{1})\cos (x_{2})+3t\sin ^{2}\cos (x_{2})  \nonumber
\\
u_{0}(\mathbf{x}) &=&0,\;u_{1}(\mathbf{x})=\sin ^{2}(x_{1})\cos (x_{2})
\end{eqnarray}
with the exact solution $u(x,t)=t\sin ^{2}(x_{1})\cos (x_{2})$ that clearly
shows what we have stated above.

At this point we want to mention a salient characteristic of the
new trend mentioned above: the authors fabricate unphysical
problems with exact solutions and solve them by means of methods
like ADM, HPM, and HAM.

\section{Further comments and Conclusions}

\label{sec:conclusions} The discussion above clearly shows that the paper of
Ko\c{c}ak and Y\i ld\i r\i m\cite{KY09} is another example of the new trend
in mathematical physics that consists of obtaining the textbook Taylor
series by means of elaborate approximation methods. We proved that this
standard approach enables one to solve their differential equations exactly.
However, those authors did not obtain the exact result even at fifth order
of HPM. The reason appears to be a wrong calculation of the correction of
first order. For example, in the introductory simple equation they obtained $%
u^{(0)}=0$ and $u^{(1)}=\Theta (t)t\delta (x)$, where $\Theta (t)$ is the
Heaviside step function, but this solution does not satisfy the differential
equation $\partial ^{2}u^{(1)}/\partial t^{2}=\delta (t)\delta (x)$. It is
wrong by a factor two and seems to be the reason why Ko\c{c}ak and Y\i ld\i r%
\i m\cite{KY09} did not obtain the exact result at second order as predicted
by the straightforward Taylor series. However, this sloppiness is weightless
when contrasted with such magnificent contribution to nowadays science and
the referees and editors of the journal are content with it.

Another issue is the physical utility of the models considered by Ko\c{c}ak
and Y\i ld\i r\i m\cite{KY09}. Notice, for example, that the amplitude of
the solution $\mathbf{u}(\mathbf{x},t)$ given by equations (20)--(23) in
that paper, which is of the form $\mathbf{u}(\mathbf{x},t)=t\mathbf{u}_{1}(%
\mathbf{x})$ discussed above in Sec.~\ref{sec:Taylor}, increases unboundedly
with time. But the new trend will not be deterred by such prosaic
considerations.

The reader may find the discussion of other articles that belong to the new
trend in mathematical physics elsewhere\cite
{F07,F08b,F08c,F08d,F08e,F08f,F09a,F09b,F09c}. We recommend the most
interesting case of the predator--prey model that predicts a negative number
of rabbits\cite{F08d}.

Finally, we mention that a slightly different version of this comment was
rejected on the basis that ``I have determined that it lacks the qualities
of significant timeliness and novelty that we are seeking in this journal.
We request you to consider submitting your manuscript in another forum that
would better suit the material your manuscript covers''. After pondering a
while we realized that we submitted our manuscript a couple of weeks after
the appearance of that paper. The new trend is advancing so fast that we
delayed rather too much.

\end{document}